\documentclass[%
 reprint, superscriptaddress,
showpacs,preprintnumbers,
notitlepage,
 amsmath,amssymb,
 aps,
prx,
floatfix,
]{revtex4-1}

\usepackage{graphicx}% Include
\usepackage{xcolor}
\usepackage{bm}% bold math
\newsavebox{\measurebox}

\usepackage[utf8]{inputenc}
\usepackage{graphicx}
\usepackage{tcolorbox}
\usepackage{dirtytalk}
\usepackage{xr}
\usepackage{soul}

\begin{document}

\title{The anatomy of social dynamics in escape rooms}

\author{Rebeka O. Szabo}
\affiliation{Department of Network and Data Science, Central European University, Budapest, Hungary}
\affiliation{Laboratory for Networks, Innovation and Technology, Corvinus University, Budapest, Hungary}

\author{Sandeep Chowdhary}
\affiliation{Department of Network and Data Science, Central European University, Vienna, Austria}

\author{David Deritei}
\affiliation{Department of Molecular Biology, Semmelweis University, Budapest, Hungary}

\author{Federico Battiston}
\affiliation{Department of Network and Data Science, Central European University, Vienna, Austria}

\begin{abstract}
From sport and science production to everyday life, higher-level pursuits demand collaboration. Despite an increase in the number of data-driven studies on human behavior, the social dynamics of collaborative problem solving are still largely unexplored with network science and other computational and quantitative tools. Here we introduce escape rooms as a non-interventional and minimally biased social laboratory, which allows us to capture at a high resolution real-time communications in small project teams. Our analysis portrays a nuanced picture of different dimensions of social dynamics. We reveal how socio-demographic characteristics impact problem solving and the importance of prior relationships for enhanced interactions. We extract key conversation rules from motif analysis, and discuss turn-usurping gendered behavior, a phenomenon particularly strong in male dominated teams. We investigate the temporal evolution of signed and group interactions, finding that a minimum level of tense communication might be beneficial for collective problem solving, and revealing differences in the behavior of successful and failed teams. Our work unveils the innovative potential of escape rooms to study teams in their complexity, contributing to a deeper understanding of the micro-dynamics of collaborative team processes.
\end{abstract}

\maketitle

\section*{Introduction}

From the viral spread of rumours to the emergence of large-scale cooperation, human societies produce social dynamics and collective endeavours often hard to understand, characterize and predict. At the heart of this phenomenon is the innate need and ability of humans to collaborate and connect to others~\cite{uzzi1999embeddedness}.
Social interactions are indeed key to understand information exchange~\cite{granovetter1973strength} and social contagion~\cite{centola2007complex, aral2012identifying, lehmann2018complex, guilbeault2018complex}. In recent years, advances in technologies have allowed us to track at unprecedented fine-grained scale face-to-face communication patterns in a variety of different contexts~\cite{pentland2005socially, eagle2006reality, pentland2008honest, callahan2016pentland, sekara2016fundamental, lederman2017open, sapiezynski2019interaction}, drawing attention to the importance of high-frequency time-resolved social processes~\cite{pentland2012new} to understand  collective intelligence~\cite{woolley2010evidence} and collaborative problem solving~\cite{de2014strength, sekara2016fundamental,gomez2019clustering, monechi2019efficient}.
 In the last decades, network science has proved to be a powerful and flexible framework to understand the complex relational structure of human dynamics~\cite{christakis2009connected, easley2010networks}, from structural balance theory~\cite{fritz1958psychology, rapoport1963mathematical} to the detection of emergent mesoscale structures such as communities~\cite{girvan2002community} and cores~\cite{borgatti2000models}, associated with the coordinated behavior of multiple individuals in human societies.

 From sport~\cite{fewell2012basketball} to science~\cite{guimera2005team}, a crucial dimension of social interactions is encoded in teams. Indeed, whenever a goal needs to be achieved, people naturally join forces to overcome the limitations imposed by the capabilities of single individuals. In science, the importance of teams is increasingly recognised, as collective efforts are becoming more and more important for the production of knowledge, not only in volume but also in impact and attention~\cite{wuchty2007increasing}. 
The predominance of team success has been attributed to the increasing need for specialized knowledge from different domains recombined through interdisciplinary collaborations to solve modern day problems~\cite{hunter2008collaborative,jones2008multi, torrisi2019creative}, highlighting the importance of team composition~\cite{jackson2003recent}. Past investigations about drivers of success in teams have revealed a complex and multifaceted picture. Diverse~\cite{guimera2005team} and fresh teams~\cite{zeng2021fresh} may help widen skills and perspectives, though potentially introduce conflict and hinder an efficient communication~\cite{uzzi2012scientific}.
Other known determinants of team success include collaborations across multiple institutions~\cite{jones2008multi}, inter-member familiarity~\cite{harrison2003time} and prior shared successes of teams~\cite{petersen2015quantifying, mukherjee2019prior}. Going beyond the simple dichotomy between solo and synergistic work, also size was found to be relevant, with larger teams developing science by focusing on more recent trends, in contrast to smaller ones, tending instead to dig deeper into past literature and often producing more innovative recombination of ideas~\cite{wu2019large}. Even in apparently non-collaborative performances, such as those of elite athletes in individual disciplines such as tennis or martial arts, the presence of organized, task-focused supportive team of people has become predominant and key for succeeding. All in all, teams serve as 'petri-dishes' for social influence and are often considered factories of innovation~\cite{borner2010multi,uzzi2013atypical}.

\begin{figure*}[!htb]
    \centering
    \begin{minipage}{1\textwidth}
        % \subfloat[]
        {\includegraphics[scale=.7, trim =1cm 0.5cm 0cm 1cm]{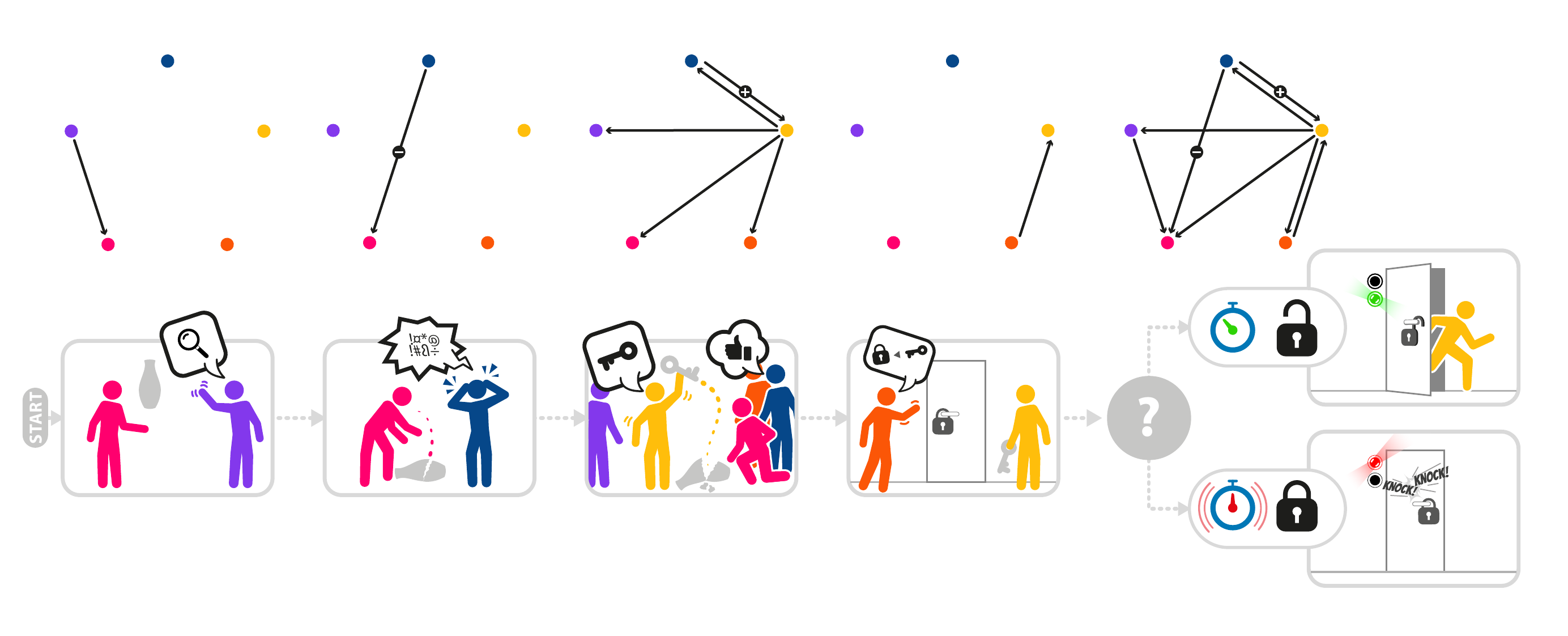}}
    \end{minipage}%
\caption{\textbf{The complex and diverse dimensions of social interactions in escape rooms.} We collect the social interactions of teams, who must collaborate in the non-routine environment of escape rooms to successfully solve tasks and thereby exit the room within an hour. Interactions are represented as temporal networks, which capture skeletal structure of communication between team members. Each interaction is directed, can be emotionally neutral or charged (either positively or negatively), addressed to a particular person (pairwise) or to a wider group.}
    \label{fig:schematic}
\end{figure*}

In this work we study team dynamics in the innovative settings of escape rooms. Escape rooms are free from the typical frailties of traditional laboratory experiments and field studies. Similarly to experiments, they provide the same controlled environment for all groups under observation, yet minimizing the potential modification of participants' behavior as a response to being examined by researchers~\cite{landsberger1958hawthorne}. Besides, social interactions can be followed and recorded at a high-frequency, allowing us to observe intact, non-manipulated teams in a nuanced, meticulous manner, overcoming the limitations associated with temporally aggregated data~\cite{leenders2016once,kozlowski2000multilevel} or self-reported questionnaires. 
Exploiting this innovative setting, we extracted from video records the real-time verbal and nonverbal communication of 40 small problem-solving teams. We analyze their high-resolution social dynamics, including collaborative network evolution and conversation rules guiding communication, exchanges of emotions and group interactions, linking them to successful team performance. Besides, we integrate such information by exploring the wider sociodemographic characteristics of team players, including gender, age and education, and their prior acquaintanceship, and meeting frequency. By capturing high-frequency real-time social interactions, our research highlights the potential of escape rooms to investigate task-performing teams in a minimally biased environment, contributing to advance the new science of teams. 
 
\section*{Results}

We study the social interactions of 40 teams collected in two distinct escape rooms. For each team (composed of 4 or 5 individuals, for a total of 171 players, all from Hungary) we extract high-resolution temporal interactions from video records (see Methods). Teams that manage to escape within one hour are deemed 'successful', while teams unable to do so are labeled as 'failed' groups. Teams must explore and exploit information, possibly through collaborations, by searching for clues, deciphering codes or opening locks. None of these tasks require any specific knowledge, skills or capabilities. However, experienced escape room players can be expected to have an advantage. Therefore, all teams in our data contain inexperienced members only (first-timers, or people who played maximum once in another room). 
 For each team, we also recorded the sociodemographic characteristics of each player (such as age, gender and education), and relational data among team members (such as prior acquaintanceship and meeting frequency),  by a questionnaire filled individually right before the game. 

Our data can be mapped as a temporal network, where interactions among individuals occur at specific points in time~\cite{holme2012temporal}. In Fig.~\ref{fig:schematic} we present a schematic picture of collaborative social activities in escape rooms and its network representation at five different temporal snapshots. 
We record both verbal and non-verbal (e.g. showing something) interaction ties between team members, between one sender and one receiver (pairwise interaction) or more than one receiver (group interaction). Interactions are directed, and assigned either a neutral, positive or negative sign (see Methods). In the following, we provide a data-driven characterization of different dimensions of the social dynamics taking place in escape rooms. 

\begin{figure*}[!htb]
    \centering
    \begin{minipage}{1\textwidth}
        % \subfloat[]
        {\includegraphics[scale=.7, trim = 0cm 12cm 0cm 1cm, clip]{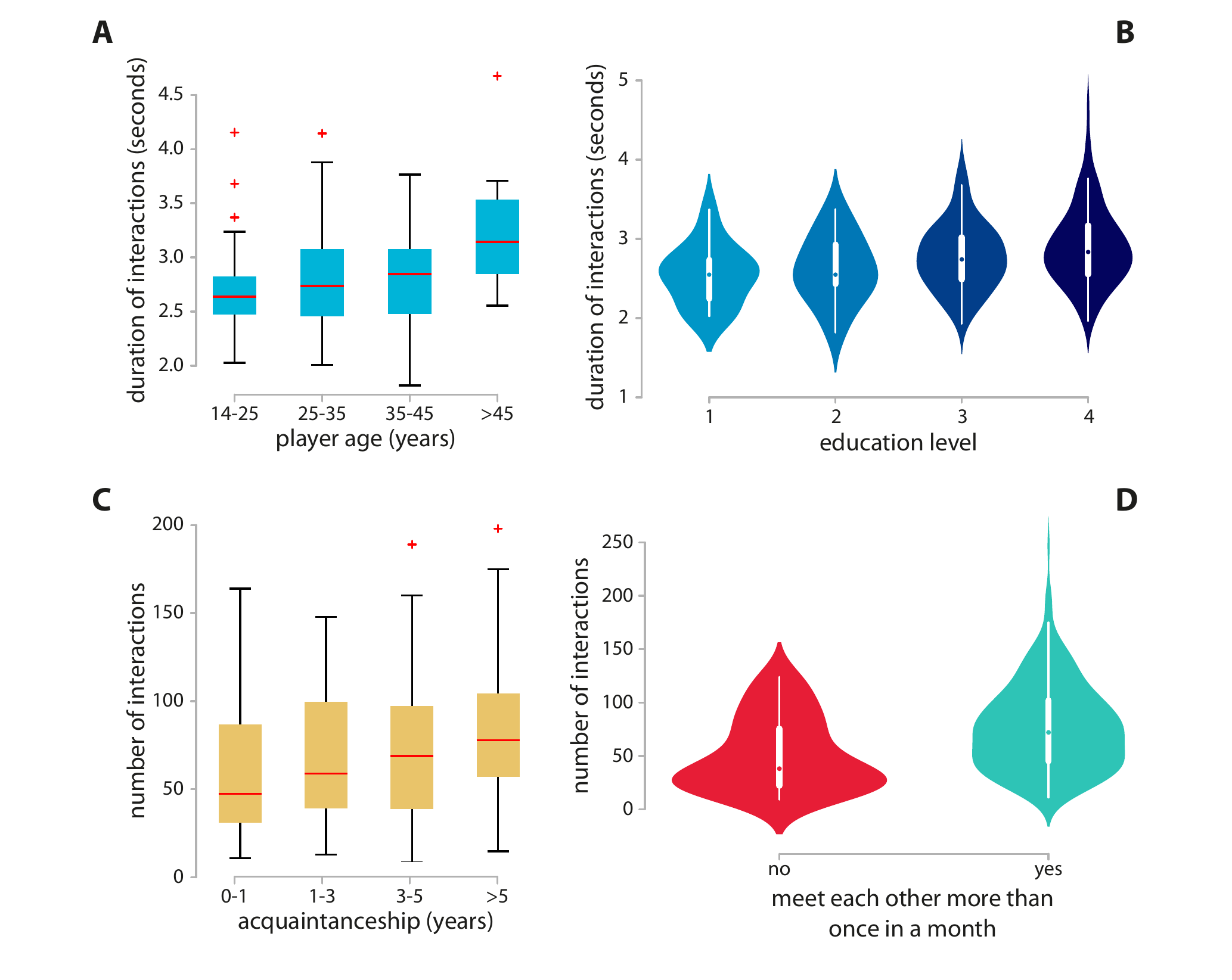}}
    \end{minipage}%

\caption{\textbf{General features of team interactions.} Individual demographics and relational characteristics of players determine their interaction patterns. Older (A) and more educated (B) players sustain longer interactions. Moreover, both the length of prior acquaintanceship (C) and meeting frequency (D) are associated with an increased number of interactions during problem-solving.}
    \label{fig:macro}
\end{figure*}

\subsection*{General features of collaborative groups}
Conversations in escape rooms are fast-paced, with teams having on average 30 interactions per minute among members, each one typically lasting 3 seconds.
We find that older and more educated players speak for longer stretches at a time (Figs.~\ref{fig:macro}A,B). The strength of a relationship between individuals has a significant role in determining the intensity of their collaborative pairwise interactions during problem-solving. In particular, prior member familiarity promotes communication, and the more time two players have known each other, the higher their rate of communications (Fig.~\ref{fig:macro}C). Moreover, people who meet frequently (more than once in a month) interact approximately 1.6 times more during problem-solving than those with lower meeting frequency (Fig.~\ref{fig:macro}D)

\subsection*{The microscopic architecture of team interactions}
\begin{figure*}[!htb]
    \centering
    \begin{minipage}{1\textwidth}
        % \subfloat[]
        {\includegraphics[scale=.7, trim = 3cm 3.5cm 0cm 1cm]{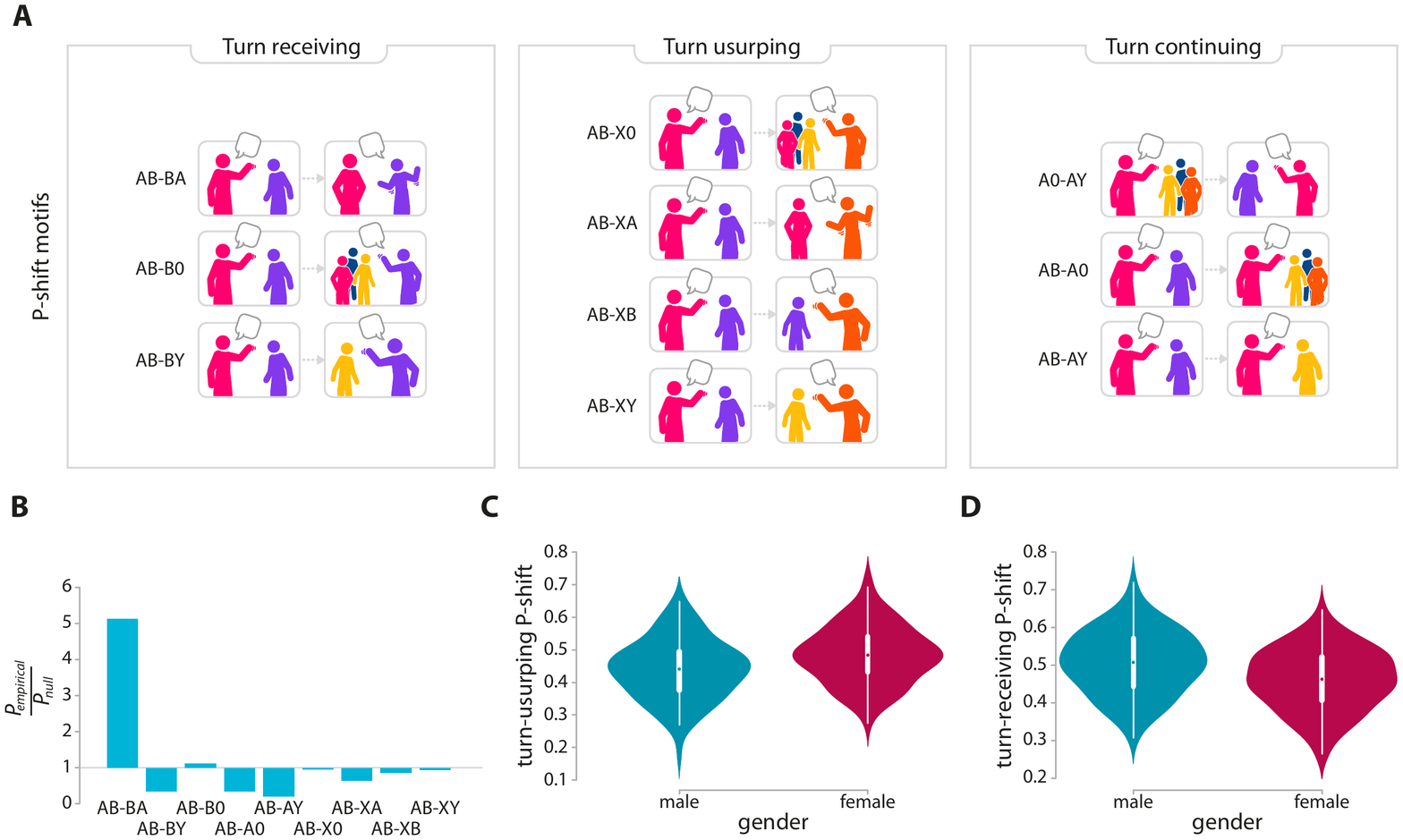}}
    \end{minipage}%

\caption{\textbf{Microscopic nature of team interactions.} We identify ordered communication sequences in escape rooms by Gibson's Participation Shift Profile framework~\cite{gibson2003participation}. P-shift motifs started with a directed remark (pairwise) can be clubbed up into 3 categories denoting the way by which the second speaker get they turn (A). The prototypical turn-receiving P-shift \textit{AB-BA} (which refers to an immediate reciprocation by \textit{B} is overrepresented at the expense of \textit{AB-BY}, while pairwise turn-continuing motifs such as \textit{AB-AY} and \textit{AB-A0} are underrepresented in escape rooms (B). Men and women participate differently in team conversations, with women showing lower levels of turn-receiving behaviour (C). This is probably connected to the observation that women are more likely to usurp turns (D) to express their opinion, as they rarely receive it. 
}
    \label{fig:micro}
\end{figure*}

Conversations involve participants taking turns to speak, punctuated by stretches of silence. Changes in these turns among the participants are governed by rules, ensuring a basic level of order and intelligibility. To understand which rules govern collaborative communication in escape rooms, we use the Participation shift profile (P-shifts) framework developed by Gibson~\cite{gibson2003participation}. 
P-shifts are behaviourally meaningful and easily interpretable inventory of network motifs, small subgroup patterns that carry information about the underlying mechanisms of social interactions.
We count the frequency of all possible P-shifts associated with pairwise interactions, capturing the rules of conversation a team adopts. P-shifts can be categorized into three types -- turn-receiving, turn-continuing and turn-usurping -- for interpretation purposes (Fig.~\ref{fig:micro}A), on which we elaborate in the following. By comparing them with what found in a suitable null model, we determine whether the motif frequencies of the empirical data are significantly different from those observed by random chance in systems which preserve the same number of interactions, but where temporal correlations are washed away (Fig.~\ref{fig:micro}B).

The most frequent P-shift is \textit{AB-BA}, a turn-receiving shift, where the second speaker \textit{B} receives the entitlement to speak from the first speaker \textit{A}. For this reason, the \textit{AB-BA} motif is often referred to as the “current-select” rule, and usually covers questions, commands, requests or accusation-denial type of action-reaction pairs. The over-representation of this motif suggests a high demand for practical actions and task implementation under time pressure in escape rooms. Another turn-receiving motif, \textit{AB-BY}, was instead significantly underrepresented in our data. \textit{AB-BY} refers to those situations when an unaddressed recipient is turned into a target in the next speaking turn. This interaction chain is rare, in contrast to \textit{AB-BA}, indicating that escape room players tend to have repeated exchanges as pairs, possibly associated with the presence of subgroups within teams. Turn continuing motifs (such as \textit{AB-A0} and \textit{AB-AY}) where the same speaker shifts from one target to another, either a group or a third person, are also found to be statistically underrepresented.

Interestingly, a few roles tend to be associated with particular socio-demographic characteristics, such as age, education and gender, highlighting -- in Gibson's words -- a \textit{role differentiation}~\cite{gibson2003participation}. 
This can be easily quantified by computing the Spearman's rank correlation $r_s$ and the Kolmogorov-Smirnov 2-sample test statistic $D$  between a specific P-shifts motif and such features. We find that older people are more inclined to address the group regardless whether they receive the turn (as \textit{B} in \textit{AB-B0}, $r_s=0.17$, $p_{val}=0.025$), or usurp the turn (as \textit{X} in \textit{AB-X0}, $r_s=0.2$, $p_{val}=0.025$).
Players with the lowest level of education (elementary school) are more inclined to assume  turn-usurper roles (both as \textit{X} in \textit{AB-XA}, $r_s=-0.2$, $p_{val}=0.007$, and in \textit{AB-XB}, $r_s=-0.16$, $p_{val}=0.037$), though they are less likely to turn-usurp by addressing the group (\textit{AB-X0}, $r_s=0.16$, $p_{val}=0.031$).

\begin{figure*}[!htb]
    \centering
    \begin{minipage}{1\textwidth}
        % \subfloat[]
        {\label{fig:signs_a}\includegraphics[scale=.7, trim = 0cm 13.5cm 0cm 1cm, clip]{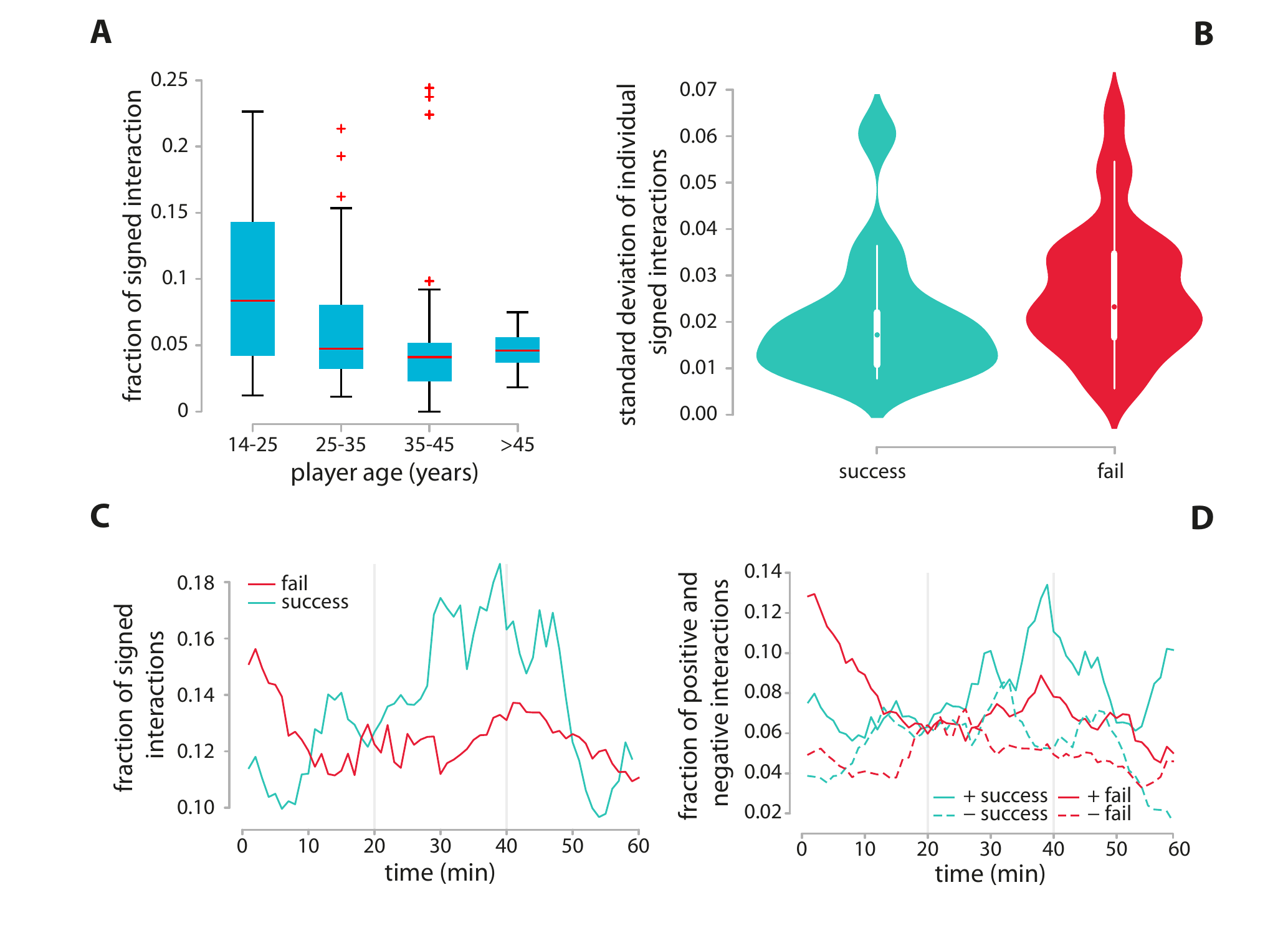}}
    \end{minipage}%

\caption{\textbf{Signed interactions.} Older players initiate less signed communication than younger members (A), suggesting a greater focus on task performance. While the total amount of emotionally charged interactions is similar in successful and failed teams, successful groups are more emotionally balanced across their individuals than the failed ones. This is quantified by the standard deviations of signed interactions across team members (B). Failed teams are characterized by a greater frequency of positive interactions in the initial stage of the game, which however rapidly declines over time (C). This trend is largely due to the high initial frequency of positive ties (D), possibly suggesting a lower amount of task focus. Surprisingly, successful teams show a slightly higher rate of negative interactions, suggesting that a minimum level of tense communication might be beneficial for collective problem solving.}
    \label{fig:signs}

\end{figure*}

We also observe correlations between gender and given roles in escape rooms. We find that women typically turn usurp (as \textit{X} in \textit{AB-X0/AB-XA/AB-XB/AB-XY}, $D=0.23$, $p_{val}=0.02$, Fig.~\ref{fig:micro}C) at the expense of receiving the turn to speak (as \textit{B} in \textit{AB-BA/AB-B0/AB-BY}, $D=0.23$, $p_{val}=0.019$, Fig.~\ref{fig:micro}D). This shows that women are rarely given the opportunity to speak after being addressed, and that they have to rather seize the floor in order to express their opinion. Indeed, these associations are even more pronounced ($D=0.61$, $p_{val}=0.039$ and $D=0.78$, $p_{val}=0.003$, respectively) when we reduced our analysis to male dominated teams only, where there is a single woman in the team.
We also observe that the lack of turn receiving opportunities lead women in male dominated groups to communicate much less than their teammates (as measured by their fraction of outgoing communication ties, $D=0.611$, $p_{val}=0.04$), a finding which is neither present in balanced groups nor in women dominated groups. This suggests that the social environment can impose constraints on women communication and relationship behavior in task-focused groups, complementing previous observations \cite{ni2021gendered}.

\begin{figure*}[!htb]
    \centering
    \begin{minipage}{1\textwidth}
        % \subfloat[]
        {\includegraphics[scale=.7, trim = 0cm 12cm 0cm 1cm, clip]{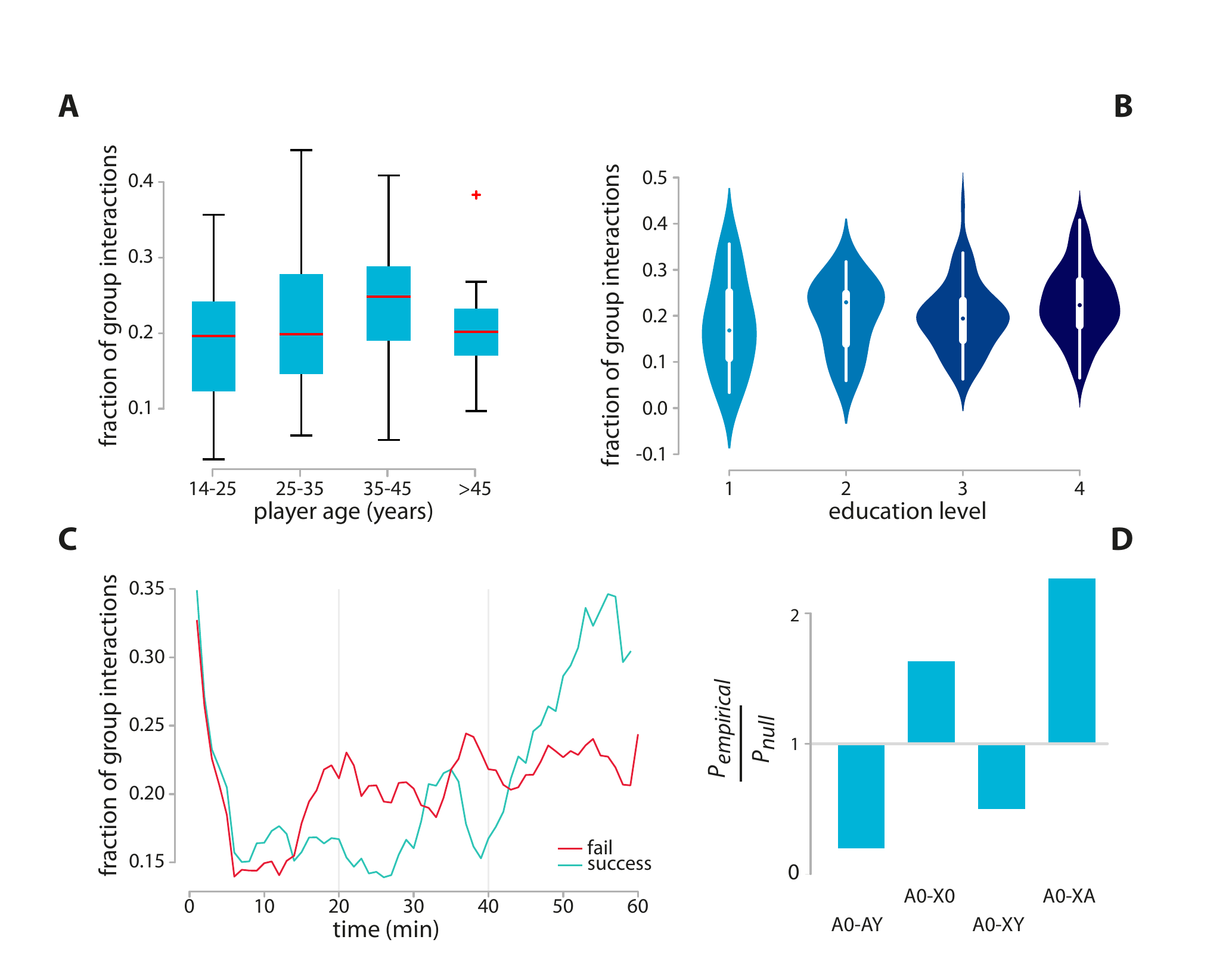}}
    \end{minipage}%

\caption{\textbf{Group interactions.} Older (A) and more educated (B) players display a higher frequency of group interactions, suggesting an unequal level of confidence and authority in the teams. An investigation of the dynamics of group interactions reveal that after an initial get-together members tend to work in smaller groups (C). An increase in group interactions for failed teams around the 20-minute mark reflects the emergence of possible early issues, while the rapid increase for successful teams in the latest part of the game is associated with celebration. P-shift analysis reveals the overabundance of motifs \textit{A0-X0} and \textit{A0-XA}, typically associated with the delivery of complex messages (D).}
    \label{fig:group}
\end{figure*}

\subsection*{Signed interactions}

Our experiment also gave us access to information about the emotional load of each interactions, classified as neutral, positive or negative. Praising, encouragement, and more in general every relationship-oriented behavior that has a positive effect on team spirit was classified as a positive tie. Negative edges were considered those that influence team spirit adversely such as creating tension, provocation or disparagement. The wide majority of the recorded task-related exchanges were classified as neutral. However, if a task-related remark is emotionally loaded (e.g. somebody is yelling), a non-neutral sign is assigned to the communication tie (negative). This classification aims to mimic the well-established distinction between task-related and relationship-related social behavior \cite{jehn1997qualitative,simons2000task,de2003task}. More details can be found in the Methods section.

We find that both older (Fig.~\ref{fig:signs}A) and more educated players tend to have a smaller amount of emotionally loaded interactions ($r_s=-0.27$, $p_{val}=0.0003$ and $r_s=-0.18$, $p_{val}=0.021$, respectively). By looking at their signs, we observe that the fraction of negative ties are the ones that account for the previous, negative correlations ($r_s=-0.3$, $p_{val}=0.00007$ and $r_s=-0.23$, $p_{val}=0.003$, for age and education level, respectively), while these two socio-demograhic features are not associated with higher chances of initiating positive interactions. These associations suggest that older and more highly educated team members are less likely to engage in social behavior that has a negative effect on team spirit. 

Although there is a similar amount of emotional interactions in both successful and failed teams, Fig.~\ref{fig:signs}B suggests that these emotional interactions are not equally distributed. Successful teams are more balanced in terms of emotional charge ($D=0.40$, $p_{val}=0.078$). In other words, members of successful teams tend to initiate a similar amount of emotional interactions with one another, while in failed groups an emotional polarization emerges, where only one or two actors display a higher number of emotionally loaded behavior. An investigation of how emotional ties evolve across time during the task performance reveals different temporal patterns for successful and failed teams (Fig.~\ref{fig:signs}C). During the initial stages of the game, failed teams show a higher rate of emotionally loaded interactions. However, over time the rate of non-neutral interactions becomes higher for successful teams.

We can gain more insights on these patterns by decomposing signed interactions into positive and negative ties (Fig.~\ref{fig:signs}D). Surprisingly, the initially greater amount of emotionally loaded interactions in failed teams
is due to a high rate of positive interactions ($13\%$, doubling the amount of successful teams), which however rapidly declines within the first 15 minutes. This unexpected feature might reflect a lower focus (e.g. making jokes, laughing), which will eventually reveal crucial for the outcome of the game. As the game progresses, we find a greater frequency of positive interactions for successful teams, reaching $14\%$ after $~40$ minutes, when many crucial tasks have already been completed. The dynamics of negative interactions is less rich. Negative ties are less frequent than positive ones for both successful and failed teams, and show no significant temporal patterns, with the exception of the rapid decline in negative interactions for successful teams towards the very end of the game. The slightly greater rate of negative ties in successful teams compared to failed ones ($6\%$ and $4\%$ respectively) suggests that a minimum level of tense communication might be beneficial for task performance in collective problem solving. As some successful teams start leaving the room around the 47 minute mark, only the remaining successful teams contribute to the curves shown in Fig.~\ref{fig:signs}C,D.

\subsection*{Group interactions}
Beyond pairwise communication, we also investigate group interactions~\cite{battiston2020networks}, where more than a single recipient is addressed. Group interactions constitute $\approx 20 \%$ of all the interactions, and are typically 0.77 shorter than one-to-one interactions. We find that older (Fig.~\ref{fig:group}A) and more educated (Fig.~\ref{fig:group}B) people have a higher frequency of group interactions ($r_s=0.23$, $p_{val}=0.003$, and $r_s=0.2$, $p_{val}=0.009$, respectively), suggesting an unequal level of confidence and authority in the teams. Observing the temporal dynamics (Fig.~\ref{fig:group}C), teams tend to start the game with a high amount of group interactions. This pattern is associated with teams ‘warming up’, familiarizing themselves with the environment, and discussing together a strategy. 
Such a number falls abruptly after the first 5 minutes, reflecting division of work and individual focus. Such temporal group dynamics also display differences between successful and failed teams. Failed teams show a greater amount of group interactions around the 20-minute mark, associated with the emergence of potential early problems. By contrast, successful teams only display a low rate of group interactions around the same time, possibly reflecting greater focus and productivity. After manual visual inspection from the videos, we were able to associate the peak in group interactions in successful teams at the 30-minute mark to the presence of productive halftime get-together, where members synthesize their knowledge, discuss achievements how to proceed. As a few successful teams managed to escape slightly before the end of the game, only the remaining successful teams contribute to the curves shown in Fig.~\ref{fig:group}C.

Conversation rules governing group interactions can be analysed by a suitable extension of the previously introduced P-shift profiles~\cite{gibson2003participation}. Following a group remark, where \textit{0} identifies the group, either the speaker continues their turn (as \textit{A} in \textit{A0-AY}, turn continuing), or someone else claims the turn (turn claiming). Turn-claiming can happen in three distinct ways, as \textit{X} in \textit{A0-XA/A0-X0/A0-XY}, where the addressee can be the first speaker \textit{A}, the group \textit{0}, or another person \textit{Y}. Observing the pattern of ordered group interactions accounting for conversation rules, we find that when someone claims the floor, it typically happens in the form of \textit{A0-X0} or \textit{A0-XA} at the expense of \textit{A0-XY} (Fig.~\ref{fig:group}D). These two P-shifts overrepresented in our data cover those situations when a remark addressed to the group is followed by a reaction.

To assess the relevance of each motif, we compare its  frequency in the real data against what observed by random chance in systems which preserve the same number of interactions but where temporal correlations have been eliminated (Fig.~\ref{fig:group}A, no significant difference between successful and failed teams). We find that the most overabundant motif is \textit{A0-XA}, the 'group version' of the previously discussed \textit{AB-BA} pairwise motif~\cite{gibson2003participation}. A second group motif, \textit{A0-X0}, is also found to be overabundant. Both of these P-shifts (someone addressing a group followed either by a second speaker addressing the group again or a team representative replying to the original communicator) are usually associated with the delivery of complex messages. For example, after \textit{A} proposes an idea or gives an instruction to the group, \textit{X} takes the floor to explain or translate \textit{A}’s idea (\textit{A0-X0}), or asks for clarification to \textit{A}’s action on behalf of the team (\textit{A0-XA}).

\raggedbottom
\section*{Discussion}
Despite recent intensive efforts in characterizing human behavior from large-scale data, understanding the social and demographic drivers of successful team interactions is still a largely open and widely debated research area. Here we provided for the first time a characterization of the social dynamics of team interactions in escape rooms, non-interventional social laboratories previously unexplored in a fine-grained quantitative manner. By capturing high-frequency real-time social interactions in this innovative quasi-experimental setting, we were able to extract the building blocks of cooperative work. Our analysis revealed that socio-demographic characteristics may impact problem-solving communication. For instance, older and highly educated actors were observed to speak for longer, more often to the group, and initiate less emotional (in particular less negative) interactions, while prior strength of relationships between group members was positively  associated with enhanced social interaction.

An investigation of P-shift profiles revealed the high demand for practical actions under time pressure, manifested in conversation rules such as the "current-select" rule (\textit{AB-BA}), associated with frequent pairwise action-reaction exchanges.
Interestingly, the behavior of men and women was found to be characterized by different conversation rules, with women
often forced to turn-usurp in order to express their opinion, a pattern particularly overabundant in teams with only one female member. Successful teams displayed a higher emotional balance across their members, possibly reflecting the higher collective nature of team organization. In spite of a tendency for prosocial communication at the initial stage of the game, a temporal analysis revealed that already after 20 minutes failed teams had different interaction patterns from successful ones, displaying less task-focused behavior and the first signs of social conflict. Interestingly, also prosperous groups  were found to maintain a non-negligible number of negative interactions until the end of the game, suggesting that a minimum level of tense communication might be beneficial for collective problem solving.

In summary, here we have proposed escape rooms as an innovative research setting for studying groups in controlled, yet non-manipulated environment, where one can obtain high resolution data on collective behavior.
By investigating social dynamics at a fine-grained scale, we were able to portray an innovative and nuanced picture of the collective actions of these project teams. In the future, we intend to further investigate the division of work in this problem-solving groups to understand how exploration and exploitation tasks impose different demands on teams to allocate their resources, and how they are associated with bottlenecks slowing down collective performance. 
We hope that these insights will spark more research on team processes in escape rooms, easily accessible social laboratories contributing to a deeper understanding of intra-group dynamics and collective action.

\section*{Methods} \label{Methods}

\textit{Data processing.} Interaction data on the problem-solving activity of teams were extracted from the video records by selected trained transcribers who were trained through several individual sessions. As a test of inter-coder reliability, following individual training all applicants received the same video to code as a trial. Transcribers were given a detailed guide on how to translate video information into a temporal signed edge list. Before data were fully analysed, consistency among and within transcribers was quantified through the Krippendorff alpha test, for which we obtained a value of 0.73, well above the conventional threshold of 0.65~\cite{krippendorff2004reliability}. After the transcribers selection was completed, randomly chosen samples of the transcriptions were double-checked in a systematic manner, obtaining similar consistency values.
 
\textit{Signed interactions.}
We categorized the emotional load of the interactions and associated to each tie either a positive, negative or neutral sign. A precise guide on the classification of signed interactions was shared with the transcribers. Typical examples of positive interactions are
praising and the encouragement of one another, while disparagement, provocation and any other interactions creating tension between group members are considered negative. Task-related interactions, no matter if they express agreement or disagreement, are classified as neutral, as soon as they do not share any of the features described above.
In short, neutral interactions cover all emotionally indifferent, task-focused communication, while signed interactions are those influencing team spirit positively/negatively, and which are constructive/disruptive to social relations in the context of a collaboration-demanding, pressed environment. Inter-coder consistency for signed interactions was evaluated as part of the consistency test described above.
A 5-minute rolling window was applied for plots showing the temporal evolution of signed interactions (Fig.~\ref{fig:signs}C,D)

\textit{Group interactions.} When three or more individuals were found to interact together, all actors involved were considered members of the same group, encoding communication beyond dyadic ties and traditional links. A detailed review on the formalism and mathematical tools used to properly encode group interactions can be found in Ref.~\cite{battiston2020networks}. The ratio between the number of group interactions and the pairwise one can be computed by dividing the total number of hyperedge of order greater than two with the number of hyperedges of order equal to two. A 5-minute rolling window was applied for plots showing the temporal evolution of  group of individuals (Fig.~\ref{fig:group}C).

\section*{Acknowledgements}
R.O.S. and F.B. wish to thank to Balázs Lengyel and László Lőrincz for their feedback, and Szabolcs Tóth-Zs for his help producing the visualizations.
R.O.S. also acknowledges useful conversations with Adam R. Pah.

\section*{References}
\bibliographystyle{ieeetr} 
\bibliography{main}

\end{document}